# Asymptotic Behavior of the β Function in the $\varphi^4$ Theory: A Scheme Without Complex Parameters

## I. M. Suslov

*Kapitza Institute for Physical Problems, Russian Academy of Sciences, ul. Kosygina 2, Moscow, 119334 Russia*
*e-mail: suslov@kapitza.ras.ru*

**Abstract**—The previously-obtained analytical asymptotic expressions for the Gell-Mann–Low function $\beta(g)$ and anomalous dimensions in the $\varphi^4$ theory in the limit $g \longrightarrow \infty$ are based on the parametric representation of the form $g = f(t)$, $\beta(g) = f_1(t)$ (where $t \propto g_0^{-1/2}$ is the running parameter related to the bare charge $g_0$), which is simplified in the complex $t$ plane near a zero of one of the functional integrals. In this work, it has been shown that the parametric representation has a singularity at $t \longrightarrow 0$; for this reason, similar results can be obtained for real $g_0$ values. The problem of the correct transition to the strong-coupling regime is simultaneously solved; in particular, the constancy of the bare or renormalized mass is not a correct condition of this transition. A partial proof has been given for the theorem of the renormalizability in the strong-coupling region.

## 1. INTRODUCTION

In recent works [1, 2], it was shown that the asymptotic behavior of the Gell-Mann–Low function $\beta(g)$ in the limit $g \longrightarrow \infty$ for actual field theories can be obtained analytically. The expression of the β function in terms of functional integrals leads to the parametric representation

$$g = f(t), \quad \beta(g) = f_1(t), \qquad (1)$$

where $t$ is the running parameter related to the bare charge $g_0$ as $t \propto g_0^{-1/2}$. The analysis of Eqs. (1) indicates that infinite $g$ values are related to a zero of one of the functional integrals; Eqs. (1) near such a zero are strongly simplified and the parametric representation is resolved in an explicit form. The asymptotic behavior of the β function in the $\varphi^4$ theory [1] and QED [2] is linear and the anomalous dimensions have constant limits.

Functional integrals definitely have zeros at complex $t$ values corresponding to complex $g_0$ values. In view of this circumstance, questions regarding the Hermiticity of the initial Hamiltonian, unitarity of the $S$ matrix, etc. arise. In our opinion that there is no subject for anxiety: Eqs. (1) were derived for real $g_0$ values (this fact ensures the correct perturbation theory) and were then analytically continued to the complex plane. Since the theory is renormalizable, the bare charge $g_0$ is excluded in all of the observables and its complex-valuedness at large $g$ is of no significance: the bare theory involves artificial constructions (such as an auxiliary lattice) and has no physical meaning.

However, scientific society has a bias against complex bare parameters, which is related with the old discussion of the Lee model [3] by Pauli, Heisenberg, etc. After work [4], the Lee model was considered as physically unsatisfactory owing to the existence of states with a negative norm ("ghost states"). Recently, Bender et al. [5, 6] showed that the Lee model is an acceptable physical theory, but the out-of-date point of view was already included in many textbooks [7, 8]. In fact, the problem of complex bare parameters is completely solved in the Bogoliubov construction of the $S$ matrix [9] (for details, see Section 8).

A more significant problem is associated with the applications of the $\varphi^4$ theory in the condensed matter physics; in this case, the bare Hamiltonian has a physical meaning and its parameters certainly cannot be complex. For this reason, the strong-coupling regime seems to be inaccessible and the asymptotic expressions obtained in [1] for renormalization-group functions seem to be physically meaningless.[1]

Below, it will be shown that the parametric representation (1) in the case of the lattice interpretation of the functional integrals has a singularity at the point $t = 0$, which ensures the strong-coupling regime $g \longrightarrow \infty$ (see Sections 2, 3). Such a possibility seems intuitively more satisfactory, because the bare charge $g_0$ tends to infinity rather than to a singular point in the complex plane; this solves the problem of the physical implementation of the strong-coupling regime in the condensed matter physics. The asymptotic behavior of the β function coincides with the expression obtained in [1], whereas the results for anomalous dimensions are somewhat different (see Section 4), but this difference has no physical meaning (see Section 7). The analysis of the topology of trajectories in the $t$ plane (see Section 5) shows that they can remain on the real axis or go to the complex plane in dependence on the choice of the lattice. From the field-theory point of view, the problem of the real or complex bare parameters is associated with the choice of the renormalization scheme [11] and is physically meaningless. For the scheme most often used in the theory of phase transitions [12], the presence of trajectories in the complex plane is inevitable and the convention of the use of complex bare parameters is particularly obvious (see Section 6).

---
[1] They are of interest even in this case, because they strongly simplify the summation of divergent series in the calculation of the critical exponents (see discussion in [10]).

The lattice expansions studied in Section 3 are similar to high-temperature series in the theory of phase transitions [13]. In the field-theory context, they were considered in [14–17], where the main attention was focused on the analysis of the relation of the renormalized quantities to the bare parameters; this analysis required the knowledge of a large number of expansion terms and the use of the approximate extrapolation methods. However, of physical interest are the relations between the renormalized quantities (according to the general philosophy of renormalizability), which are simpler (see Section 4). Moreover, the problem of the correct transition to the strong-coupling regime remained to be unsolved; in particular, neither constancy of bare mass nor the constancy of the renormalized mass [15] is a correct condition for this limiting transition (see Section 4). Note that the indication to the linear asymptotic behavior of the β function was obtained as early as in [16]; recently, Frasca [18] showed that the linear asymptotic behavior is in agreement with a particular solution of the Dyson–Schwinger equations [18].

## 2. PRELIMINARY DISCUSSION

Let us consider the $n$-component $\varphi^4$ theory with the action

$$S\{\varphi\} = \int d^d x \left\{ \frac{1}{2} \sum_{\alpha=1}^{n} (\nabla \varphi_\alpha)^2 \right.$$
$$\left. + \frac{1}{2} m_0^2 \sum_{\alpha=1}^{n} \varphi_\alpha^2 + \frac{1}{4} u_0 \left( \sum_{\alpha=1}^{n} \varphi_\alpha^2 \right)^2 \right\}, \quad (2)$$
$$u_0 = g_0 \Lambda^\epsilon, \quad \epsilon = 4 - d,$$

where $g_0$ and $m_0$ are the bare charge and mass, respectively; $d$ is the dimension of space;[2] and $\Lambda$ is the momentum cutoff parameter. The most general functional integral of this theory contains $M$ multipliers of the field $\varphi$ in the pre-exponential factor,

$$Z^{(M)}_{\alpha_1...\alpha_M}(x_1, ..., x_M) = \int D\varphi \varphi_{\alpha_1}(x_1) \varphi_{\alpha_2}(x_2) \qquad (3)$$
$$... \varphi_{\alpha_M}(x_M) \exp(-S\{\varphi\}),$$

and will be denoted as $K_M\{p_i\}$ after transition to the momentum representation and extraction of δ-functional factors

$$Z^{(M)}_{\alpha_1...\alpha_M}(p_1, ..., p_M) \qquad (4)$$
$$= K_M\{p_i\} \mathcal{N} \delta_{p_1+...+p_M} I_{\alpha_1...\alpha_M},$$

where $I_{\alpha_1...\alpha_M}$ is the sum of the terms $\delta_{\alpha_1\alpha_2}\delta_{\alpha_3\alpha_4}...$ with all the possible pairings, and $\mathcal{N}$ is the number of the sites of the lattice on which the functional integral is defined. The integrals $K_M\{p_i\}$ are usually estimated at zero momenta and only one integral $K_2(p)$ is required at small $p$ values,

$$K_2(p) = K_2 - \tilde{K}_2 p^2 + .... \qquad (5)$$

Introducing the vertices $\Gamma^{(L, N)}$ with $N$ external lines of the field $\varphi$ and $L$ external lines of the interaction[3]; taking into account their multiplicative renormalizability [20],

$$\Gamma^{(L, N)}(p_i; g_0, m_0, \Lambda)$$
$$= Z^{N/2} \left( \frac{Z_2}{Z} \right)^{-L} \Gamma_R^{(L, N)}(p_i; g, m), \qquad (6)$$

where $g$ and $m$ are the renormalized charge and mass, respectively; and using the renormalization conditions on zero momenta,

$$\Gamma_R^{(0, 2)}(p; g, m)\big|_{p \to 0} = m^2 + p^2 + O(p^4),$$
$$\Gamma_R^{(0, 4)}(p_i; g, m)\big|_{p_i = 0} = gm^\epsilon, \qquad (7)$$
$$\Gamma_R^{(1, 2)}(p_i; g, m)\big|_{p_i = 0} = 1;$$

the Gell-Mann–Low function $\beta(g)$ and anomalous dimensions $\eta(g)$ and $\eta_2(g)$ can be determined as

$$\beta(g) = \frac{dg}{d\ln m}\bigg|_{g_0, \Lambda = \text{const}},$$
$$\eta(g) = \frac{d\ln Z}{d\ln m}\bigg|_{g_0, \Lambda = \text{const}}, \qquad (8)$$

---

[2] It is assumed that $d \leq 4$; under this condition, the $\varphi^4$ theory is renormalizable; the parameter $\epsilon$ is not assumed to be small, if the opposite assumption is not mentioned.

[3] In the diagrammatic technique where the interaction is marked by dashed lines [19].

$$\eta_2(g) = \left.\frac{d\ln Z_2}{d\ln m}\right|_{g_0,\,\Lambda\,=\,\text{const}}.$$

Expressing these functions in terms of the functional integrals, the parametric representation for them can be obtained in the form [1]

$$g = -\left(\frac{K_2}{\tilde{K}_2}\right)^{d/2}\frac{K_4 K_0}{K_2^2}, \tag{9}$$

$$\beta(g) = -\left(\frac{K_2}{\tilde{K}_2}\right)^{d/2}\frac{K_4 K_0}{K_2^2}$$
$$\times\left\{d + 2\frac{K_4'/K_4 + K_0'/K_0 - 2K_2'/K_2}{K_2'/K_2 - \tilde{K}_2'/\tilde{K}_2}\right\}, \tag{10}$$

$$\eta(g) = -\frac{2K_2\tilde{K}_2}{K_2\tilde{K}_2' - K_2'\tilde{K}_2}\left[2\frac{K_2'}{K_2} - \frac{K_0'}{K_0} - \frac{\tilde{K}_2'}{\tilde{K}_2}\right], \tag{11}$$

$$\eta_2(g) = \frac{2K_2\tilde{K}_2}{K_2\tilde{K}_2' - K_2'\tilde{K}_2}$$
$$\times\left\{\frac{K_0''K_2 - K_0 K_2''}{K_0'K_2 - K_0 K_2'} - 2\frac{K_2'}{K_2}\right\}, \tag{12}$$

where primes stand for the derivatives with respect to $m_0^2$. The right-hand sides of Eqs. (9)–(12) depend on three parameters $g_0$, $m_0$, and $\Lambda$; one of them can be taken as a running parameter; if it is expressed in terms of $g$ using Eq. (9) and is excluded from Eqs. (10)–(12), the dependence on other two parameters disappears according to general theorems [20].

According to Eq. (9), large $g$ values can be reached near the zero of $K_2$ or $\tilde{K}_2$. In the limit $\tilde{K}_2 \longrightarrow 0$, the right-hand sides of Eqs. (10)–(12) are strongly simplified and the parametric representation is resolved in the form [1]

$$\beta(g) = dg,\ \ \eta(g) = 2,\ \ \eta_2(g) = 0\ \ (g \longrightarrow \infty). \tag{13}$$

Similar results are obtained in the limit $K_2 \longrightarrow 0$, but this limit is unphysical, because it does not ensure a continuous transition to the four-dimensional case owing to the absence of divergence in Eq. (9) at $d = 4$.

The relation of the strong-coupling limit $g \longrightarrow \infty$ to the zero of one of the functional integrals was found in [1] from an analogy with the zero-dimensional case.

This relation can be derived in a more rigorous manner. After the discretization of the space, Eq. (2) is written in the form of the lattice sum

$$S\{\varphi\} = \frac{1}{2}a^d\sum_{\mathbf{x},\mathbf{x}'}J_{\mathbf{x}-\mathbf{x}'}\varphi_{\mathbf{x}}\varphi_{\mathbf{x}'}$$
$$+ \frac{1}{2}m_0^2 a^d\sum_{\mathbf{x}}\varphi_{\mathbf{x}}^2 + \frac{1}{4}g_0 a^{2d-4}\sum_{\mathbf{x}}\varphi_{\mathbf{x}}^4, \tag{14}$$

where the case $n = 1$ is considered for simplicity and $\Lambda = a^{-1}$ ($a$ is the lattice constant) is accepted. With the change

$$\varphi \longrightarrow \varphi(g_0 a^{2d-4}/4)^{-1/4} \tag{15}$$

and the parameter

$$t = (1/g_0)^{1/2}, \tag{16}$$

functional integral (3) is represented in the form

$$Z^{(M)}\{\mathbf{x}_i\} = (2t)^{\frac{\mathcal{N}+M}{2}}\int\left(\prod_{\mathbf{x}}d\varphi_{\mathbf{x}}\right)\varphi_{\mathbf{x}_1}\ldots\varphi_{\mathbf{x}_M}$$
$$\times \exp\left\{-t\sum_{\mathbf{x},\mathbf{x}'}J_{\mathbf{x}-\mathbf{x}'}\varphi_{\mathbf{x}}\varphi_{\mathbf{x}'}\right. \tag{17}$$
$$\left. - tm_0^2\sum_{\mathbf{x}}\varphi_{\mathbf{x}}^2 - \sum_{\mathbf{x}}\varphi_{\mathbf{x}}^4\right\}.$$

Hereinafter, $a = 1$ is accepted and $J_{\mathbf{x}-\mathbf{x}'}$ and $m_0^2$ are measured in the units of $\Lambda^2$. At a finite number of integrations $\mathcal{N}$, the integral converges for all $t$ values and is regular throughout the finite part of the complex $t$ plane. Singularities can appear only in the limit $\mathcal{N} \longrightarrow \infty$, but this occurs only at the points of phase transitions, when $m^2 = 0$ and the correlation radius $\xi$ is infinite; in this case, the transition to the infinite volume of the system, which is singular, is really necessary. If $m^2 \neq 0$, then owing to finiteness of the correlation radius $\xi$, the size of the system $\mathcal{L}$ can be chosen large but finite; under the condition

$$\mathcal{L} \gg \xi \gg a, \tag{18}$$

the functional integral is well approximated by its finite-dimensional analog. In the case under consideration, $m^2$ is certainly finite (in fact, $m^2 \longrightarrow \infty$ in the limit $g \longrightarrow \infty$) and the passage to the limit $\mathcal{N} \longrightarrow \infty$ is not required. For this reason, the integrals $K_M$ and their derivatives are regular in the complex $t$ plane and the appearance of infinities on the right-hand sides of Eqs. (9)–(12) can be attributed only to the zeros of denominators; in particular, they appear in Eq. (9) only near the zero of $K_2$ or $\tilde{K}_2$. The parameter $t$ is considered below as the running parameter of the parametric representation.

The transition from Eq. (2) to Eq. (14) was based on the correspondence

$$-\varphi(x)\nabla^2\varphi(x) = \varphi(x)\hat{p}^2\varphi(x) \longrightarrow \varphi_{\mathbf{x}}\epsilon(\hat{p})\varphi_{\mathbf{x}}$$
$$= \sum_{\mathbf{x}'} J_{\mathbf{x}-\mathbf{x}'}\varphi_{\mathbf{x}}\varphi_{\mathbf{x}'}, \quad (19)$$

where $\hat{p}$ is the momentum operator,

$$\epsilon(p) = \sum_{\mathbf{x}} J_{\mathbf{x}} e^{i\mathbf{p}\cdot\mathbf{x}} = \epsilon(0) + p^2 + O(p^4), \quad (20)$$

is the bare spectrum, and it is taken into account that $\exp\{i\hat{\mathbf{p}}\cdot\mathbf{x}\}$ is the operator of the translation by the vector $\mathbf{x}$. The overlap integrals $J_{\mathbf{x}}$ are assumed to be rapidly decreasing with an increase in $|\mathbf{x}|$ so that spectrum (20) is regular, $J_0 = 0$ is accepted, and the coefficient of the term $p^2$ is taken to be unity for the correspondence with the continual limit (see Eq. (19)) taking into account that $\epsilon(0)$ can be included in the renormalization of $m_0^2$.

Below we analyze the singularity of the parametric representation (9)–(12), which has a simple origin. For $g_0 \gg 1$ in Eq. (17), the expansion in the gradient term $t J_{\mathbf{x}-\mathbf{x}'}\varphi_{\mathbf{x}}\varphi_{\mathbf{x}'}$ is possible. In the zeroth order in $t$, the integral $Z^{(2)}$ in the coordinate representation has the form of a $\delta$ function, $Z^{(2)}(\mathbf{x},\mathbf{x}') \sim \delta_{\mathbf{xx}'}$ and its Fourier transform is independent of the momentum; this dependence appears only in the first order in $t$. For this reason, the integral $\tilde{K}_2$ in expansion (5) has an additional smallness as compared to $K_2$; i.e., $K_2/\tilde{K}_2 \sim 1/t$; this leads to a singularity in Eqs. (9) and (10). This singularity is more complicated than that near the zero of $\tilde{K}_2$ in the complex plane, when the other integrals and their derivatives remain finite [1]. In the limit $t \longrightarrow 0$, the integral $K_M$ has an additional factor proportional to $t^{M/2}$ as compared to $K_0$ (see Eq. (17)) and the differentiation with respect to $m_0^2$ provides the factor $t$; for this reason, the singularity in Eqs. (9)–(12) requires a more accurate analysis.

## 3. LATTICE EXPANSIONS

The expansion of functional integral (17) in the gradient term $tJ_{\mathbf{x}-\mathbf{x}'}\varphi_{\mathbf{x}}\varphi_{\mathbf{x}'}$ contains the mean values of the products of the fields $\varphi_{\mathbf{x}_1}\varphi_{\mathbf{x}_2}\ldots$ over the distribution[5]

$$P\{\varphi\} \sim \prod_{\mathbf{x}} \exp\{-tm_0^2\varphi_{\mathbf{x}}^2 - \varphi_{\mathbf{x}}^4\}. \quad (21)$$

Such mean values are nonzero only for partially (or completely) coinciding coordinates and are represented through the scheme

$$\langle \varphi_{\mathbf{x}_1}\varphi_{\mathbf{x}_2}\varphi_{\mathbf{x}_3}\varphi_{\mathbf{x}_4}\rangle = \langle\varphi_{\mathbf{x}_1}^2\rangle\langle\varphi_{\mathbf{x}_3}^2\rangle\delta_{\mathbf{x}_1\mathbf{x}_2}\delta_{\mathbf{x}_3\mathbf{x}_4}(1-\delta_{\mathbf{x}_1\mathbf{x}_3})$$
$$+ \langle\varphi_{\mathbf{x}_1}^2\rangle\langle\varphi_{\mathbf{x}_2}^2\rangle\delta_{\mathbf{x}_1\mathbf{x}_3}\delta_{\mathbf{x}_2\mathbf{x}_4}(1-\delta_{\mathbf{x}_1\mathbf{x}_2})$$
$$+ \langle\varphi_{\mathbf{x}_1}^2\rangle\langle\varphi_{\mathbf{x}_2}^2\rangle\delta_{\mathbf{x}_1\mathbf{x}_4}\delta_{\mathbf{x}_2\mathbf{x}_3}(1-\delta_{\mathbf{x}_1\mathbf{x}_2}) \quad (22)$$
$$+ \langle\varphi_{\mathbf{x}_1}^4\rangle\delta_{\mathbf{x}_1\mathbf{x}_2}\delta_{\mathbf{x}_1\mathbf{x}_3}\delta_{\mathbf{x}_1\mathbf{x}_4}$$
$$= \langle\varphi^2\rangle\langle\varphi^2\rangle(\delta_{\mathbf{x}_1\mathbf{x}_2}\delta_{\mathbf{x}_3\mathbf{x}_4} + \delta_{\mathbf{x}_1\mathbf{x}_3}\delta_{\mathbf{x}_2\mathbf{x}_4} + \delta_{\mathbf{x}_1\mathbf{x}_4}\delta_{\mathbf{x}_2\mathbf{x}_3})$$
$$+ [\langle\varphi^4\rangle - 3\langle\varphi^2\rangle\langle\varphi^2\rangle]\delta_{\mathbf{x}_1\mathbf{x}_2}\delta_{\mathbf{x}_1\mathbf{x}_3}\delta_{\mathbf{x}_1\mathbf{x}_4},$$

where the means $\langle\varphi^{2k}\rangle$ are given by the expression

$$\langle\varphi^{2k}\rangle = \frac{I_{2k}}{I_0},$$
$$I_{2k} = \int_{-\infty}^{\infty} d\varphi\, \varphi^{2k}\exp\{-tm_0^2\varphi^2 - \varphi^4\}. \quad (23)$$

Following this scheme, it is easily to obtain

$$Z^{(0)} = (2t)^{\mathcal{N}/2} I_0^{\mathcal{N}}\left[1 - t\mathcal{N}\frac{I_2}{I_0}J_0 + \ldots\right],$$

$$Z^{(2)}(\mathbf{x}_1\mathbf{x}_2) = (2t)^{\mathcal{N}/2} I_0^{\mathcal{N}}$$
$$\times 2t\left\{\frac{I_2}{I_0}\delta_{\mathbf{x}_1\mathbf{x}_2} - 2t\frac{I_2^2}{I_0^2}J_{\mathbf{x}_1-\mathbf{x}_2} + \ldots\right\},$$

$$Z^{(4)}(\mathbf{x}_1\ldots\mathbf{x}_4) = (2t)^{\mathcal{N}/2} I_0^{\mathcal{N}} (2t)^2 \quad (24)$$
$$\times \left\{\frac{I_2^2}{I_0^2}(\delta_{\mathbf{x}_1\mathbf{x}_2}\delta_{\mathbf{x}_3\mathbf{x}_4} + \delta_{\mathbf{x}_1\mathbf{x}_3}\delta_{\mathbf{x}_2\mathbf{x}_4} + \delta_{\mathbf{x}_1\mathbf{x}_4}\delta_{\mathbf{x}_2\mathbf{x}_3})\right.$$

---

[5] For technical reasons, it is convenient to retain the term $\sim tm_0^2$ in the exponential. This is not an excess of the accuracy, because $m_0^2$ is an independent parameter and $tm_0^2$ is not necessarily small at $t \ll 1$; the term $\sim tm_0^2$ in the exponential expands the region of the applicability of the expansions; this fact will be substantial below (see Section 4).

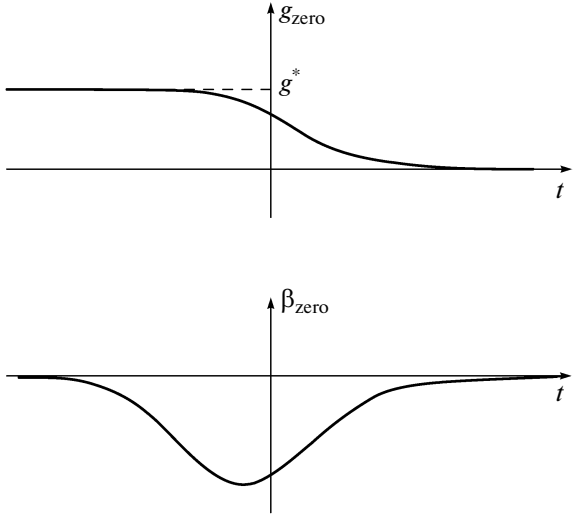

**Fig. 1.** Qualitative behavior of the functions $g_{zero}(t)$ and $\beta_{zero}(t)$ corresponding to the zero-dimensional case [1].

$$+ \left(\frac{I_4}{I_0} - 3\frac{I_2^2}{I_0^2}\right)\delta_{x_1 x_2}\delta_{x_1 x_3}\delta_{x_1 x_4} + \ldots\Bigg\}.$$

The correction term $\sim \mathcal{N} t$ to $Z^{(0)}$ disappears in view of the accepted condition $J_0 = 0$. In the general case, the terms containing $\mathcal{N}$ formally appear in the expansion, but correspond to "unconnected diagrams," when the factors $\varphi_x$ appearing from the expansion of the exponential in Eq. (17) are averaged independently of the multipliers $\varphi_{x_1}\ldots\varphi_{x_M}$ in the pre-exponential factor. It is easy to understand that the contribution of the unconnected diagrams is factorized in all of the quantities $Z^{(M)}$ in the form of the same factor and disappears in the ratios of the functional integrals. Therefore, expansions (24) are valid under the condition $t \ll 1$, and not the more strong condition $\mathcal{N} t \ll 1$.

After the transition to the momentum representation and the introduction of $K_M$ according to Eq. (4), the formulas

$$K_0 = (2t)^{\mathcal{N}/2} I_0^{\mathcal{N}},$$

$$K_2(p) = (2t)^{\mathcal{N}/2} I_0^{\mathcal{N}} \times 2t\left\{\frac{I_2}{I_0} - 2t\frac{I_2^2}{I_0^2}\epsilon(p)\right\},$$

$$K_4\{p_i\} = (2t)^{\mathcal{N}/2} I_0^{\mathcal{N}} \times (2t)^2 \frac{1}{3}\Bigg\{\left(\frac{I_4}{I_0} - 3\frac{I_2^2}{I_0^2}\right)$$

$$+ \frac{I_2^2}{I_0^2}\mathcal{N}(\delta_{p_1+p_2} + \delta_{p_1+p_3} + \delta_{p_1+p_4})\Bigg\} \quad (25)$$

are obtained. The limit $p_i \longrightarrow 0$ in $K_4\{p_i\}$ is taken according to the agreement accepted in [1]: $p_i \sim \mu$ is taken in such way as to exclude the special equalities such as $p_1 + p_2 = 0$ and then $\mu$ tends to zero. Using expansion (20) for $\epsilon(p)$ and taking into account definition (5) for $K_2$ and $\tilde{K}_2$, it is possible to arrive at the expressions

$$\frac{K_2}{K_0} = 2t\frac{I_2}{I_0}, \quad \frac{\tilde{K}_2}{K_0} = (2t)^2\frac{I_2^2}{I_0^2},$$

$$\frac{K_4}{K_0} = (2t)^2\left(\frac{I_4}{3I_0} - \frac{I_2^2}{I_0^2}\right). \quad (26)$$

Similar calculations in the $n$-component case give

$$\frac{K_2}{K_0} = 2t\frac{I_2}{nI_0}, \quad \frac{\tilde{K}_2}{K_0} = (2t)^2\frac{I_2^2}{nI_0^2},$$

$$\frac{K_4}{K_0} = (2t)^2\frac{1}{n^2}\left(\frac{n}{n+2}\frac{I_4}{I_0} - \frac{I_2^2}{I_0^2}\right), \quad (27)$$

where

$$I_{2k} = \int_0^\infty d\varphi\, \varphi^{n-1+2k}\exp\{-tm_0^2\varphi^2 - \varphi^4\}. \quad (28)$$

The representation of Eqs. (10)–(12) in the form

$$\beta(g) = -\left(\frac{K_2}{\tilde{K}_2}\right)^{d/2}\frac{K_4 K_0}{K_2^2}\left\{d + 2\frac{(\ln K_4 K_0/K_2^2)'}{(\ln K_2/\tilde{K}_2)'}\right\},$$

$$\eta(g) = 2\frac{(\ln K_2/K_0)' + (\ln K_2/\tilde{K}_2)'}{(\ln K_2/\tilde{K}_2)'}, \quad (29)$$

$$\eta_2(g) = -2\frac{(\ln K_0/K_2)'' + [(\ln K_0/K_2)']^2}{(\ln K_2/\tilde{K}_2)'(\ln K_0/K_2)'}$$

and the differentiation with respect to $m_0^2$ taking into account the relation

$$I'_{2k} = -t I_{2k+2}, \quad (30)$$

provide

$$g = \left(\frac{n I_0}{2t I_2}\right)^{d/2}\left(1 - \frac{n}{n+2}\frac{I_4 I_0}{I_2^2}\right),$$

$$\frac{\beta(g)}{g} = d + 2\frac{\dfrac{I_6 I_2}{I_0^2} - \dfrac{2 I_4^2}{I_0^2} + \dfrac{I_2^2 I_4}{I_0^3}}{\left(\dfrac{I_4}{I_0} - \dfrac{n+2}{n}\dfrac{I_2^2}{I_0^2}\right)\left(\dfrac{I_2^2}{I_0^2} - \dfrac{I_4}{I_0}\right)}, \quad (31)$$

$$\eta_2(g) = 2\frac{\dfrac{I_6 I_2}{I_0^2} - \dfrac{2 I_4^2}{I_0^2} + \dfrac{I_2^2 I_4}{I_0^3}}{\left(\dfrac{I_2^2}{I_0^2} - \dfrac{I_4}{I_0}\right)^2}, \quad \eta(g) = 0.$$

The results can be represented in a compact form by introducing the functions $g_{zero}(t)$ and $\beta_{zero}(t)$ corresponding to the zero-dimensional case [1], which have the form shown in Fig. 1:

$$g = \left(\frac{n}{2t}\frac{I_0}{I_2}\right)^{d/2} g_{zero}(tm_0^2),$$

$$\beta(g) = \left(\frac{n}{2t}\frac{I_0}{I_2}\right)^{d/2}[dg_{zero}(tm_0^2) + \beta_{zero}(tm_0^2)], \quad (32)$$

$$\eta_2(g) = \frac{\beta_{zero}(tm_0^2)}{g^* - g_{zero}(tm_0^2)}, \quad \eta(g) = 0,$$

where $g^* = 2/(n+2)$. The solution of the parametric representation in the limit $t \longrightarrow 0$ provides the asymptotic expressions

$$\beta(g) = \left[d + \frac{\beta_{zero}(0)}{g_{zero}(0)}\right]g,$$

$$\eta_2(g) = \frac{\beta_{zero}(0)}{g^* - g_{zero}(0)}, \quad \eta(g) = 0 \quad (g \longrightarrow \infty), \quad (33)$$

which are similar to Eqs. (13); i.e., the β function has a linear asymptotic behavior and the anomalous dimensions approach constant limits. The substitution of the numerical values for $n = 1$ and $d = 4$ gives $\beta(g) = 2.29g$ in agreement with [16].

However, result (33) is not final. Indeed, instead of the limit $t \longrightarrow 0$ at a constant bare mass, the limiting transition under condition $tm_0^2 \longrightarrow$ const can be considered; in this case, the structure of the theory remains unchanged, but the asymptotic behavior is different. The problem of the correct transition of the strong-coupling regime arises.

## 4. INTERPRETATION OF THE FUNCTIONAL INTEGRALS

In the framework of the diagrammatic technique, the relation of the renormalized mass to the bare mass is determined by the expansion

$$Z^{-1}m^2 = m_0^2 + u_0\frac{n+2}{2}\int\frac{d^d k}{(2\pi)^d}\frac{1}{k^2 + m^2} + \ldots, \quad (34)$$

where the contribution of the $N$th order in $u_0$ has the dimension $k^{2-\epsilon N}$ in the momentum; at $d > 2$, it is determined by the upper limit and is of the order $\Lambda^{2-\epsilon N}$; taking into account the relation $u_0 = g_0\Lambda^\epsilon$ (see Eq. (2)), the expansion

$$Z^{-1}m^2 = m_0^2 + \Lambda^2(A_1 g_0 + A_2 g_0^2 + A_3 g_0^3 + \ldots) + O(m^2 g_0(\Lambda/m)^\epsilon) \quad (35)$$

is obtained. It is usually accepted that

$$m_0^2 = m_c^2 + \delta m_0^2, \quad (36)$$

where $m_c^2$ is determined from the condition $m^2 = 0$ and, at $g_0 \ll 1$, is determined by the first term of expansion (35), i.e., $m_c^2 = -A_1 g_0 \Lambda^2$. Hence, the continual limit $\Lambda \longrightarrow \infty$ can be taken under the condition

$$m^2 = \text{const}, \quad -m_0^2 \sim g_0 \Lambda^2 \longrightarrow \infty. \quad (37)$$

By analogy, the same condition was accepted when analyzing the strong-coupling limit $g_0 \longrightarrow \infty$ [15, 21]; this procedure is based on the correct intuitive idea that the passage to the limit at $m_0^2 = $ const "is not too good," but has obvious demerits.

(i) The dependence $m_c^2 \propto g_0$ is groundlessly extrapolated to the strong-coupling region.

(ii) The equation $m^2 = 0$ can have no solutions in the strong-coupling regime (see Eq. (39)); correspondingly, decomposition (36) may be senseless.

(iii) The condition of the constant renormalized mass is not physically motivated: if particles have a finite mass $m$ in the weak-coupling region, this mass due to renormalization can both increase and decrease with growth of interaction.

Researchers usually recognized the defectiveness of such procedure and the passage to the limit under condition (37) was called the "Ising limit" [21]. According to the above discussion, the problem of the correct transition to the strong-coupling region is really unsolved.

The indicated problem in the framework of the parametric representation (9)–(12) has another meaning. The deficiency of the results such as Eqs. (32) is that the right-hand sides of the formulas depend on two independent parameters $t$ and $tm_0^2$; when one of them is excluded in favor of $g$, the dependence on $m_0^2$ remains although it should be absent according to the general theorems [20]. The problem of resolving this contradiction arises.[6]

---

[6] The dependence on $m_0^2$ for the result given by Eqs. (32) can be excluded by changing $tm_0^2 \longrightarrow t$ and multiplying $g$ and $\beta(g)$ by the same factor. However, this does not solve the problem in the general case: there are regular corrections to Eqs. (32) in $t$ that contain functions of $tm_0^2$ as coefficients.

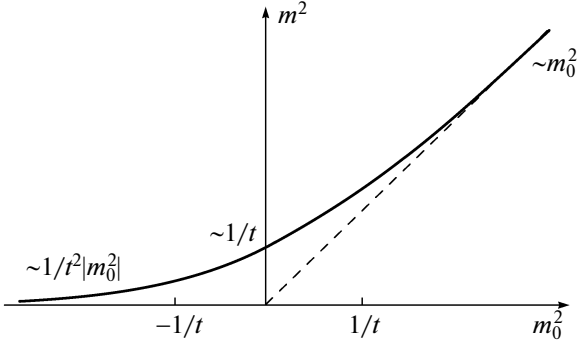

**Fig. 2.** Renormalized mass squared versus the bare mass squared in the strong-coupling region.

In reality, there is no contradiction, because the general theorems imply the appropriate passage to the continual limit $\Lambda \longrightarrow \infty$. This physically means the condition

$$m^2 \ll \Lambda^2, \qquad (38)$$

which is equivalent to the condition $\xi \gg a$ for the correlation radius; in this case, the lattice has many sites at the characteristic scale of the field variation and its further division has no significance.

In the framework of the expansions given in Section 3, the renormalized mass satisfies the relation (see [1, Eq. (53)] and Fig. 2)

$$m^2 = \frac{K_2}{\tilde{K}_2} = \frac{n}{2t} \frac{I_0}{I_2}$$
$$= \begin{cases} m_0^2, & tm_0^2 \gg 1 \\ \sim 1/t, & |tm_0^2| \lesssim 1 \\ \sim 1/t^2 |m_0^2|, & -tm_0^2 \gg 1 \end{cases} \qquad (39)$$

according to which the condition $m^2 \ll 1$ (corresponding to Eq. (38) in the dimensional units) requires the relation

$$tm_0^2 = -\kappa, \quad \kappa \gg 1. \qquad (40)$$

The change $\varphi_x^2 \longrightarrow \kappa \varphi_x^2/2$ transforms the exponential in Eq. (17) to the form

$$\exp\left\{-\frac{1}{2}t\kappa \sum_{x,x'} J_{x-x'}\varphi_x\varphi_{x'}\right\}$$
$$\times \prod_x \exp\left\{\frac{1}{4}\kappa^2(2\varphi_x^2 - \varphi_x^4)\right\}, \qquad (41)$$

where the last factor is localized near $\varphi_x^2 = 1$ and can be changed to $A\delta(\varphi_x^2 - 1)$; the constant $A$ is insignificant, because it disappears in the ratio of two integrals; hence, $A = 1$ can be accepted. As a result, Eq. (17) is modified to the form

$$Z_M\{\mathbf{x}_i\} = (t\kappa)^{\frac{N+M}{2}} \int \left(\prod_x d\varphi_x\right) \varphi_{x_1}\dots\varphi_{x_M}$$
$$\times \exp\left\{-\frac{1}{2}t\kappa\sum_{x,x'}J_{x-x'}\varphi_x\varphi_{x'}\right\}\prod_x \delta(\varphi_x^2 - 1) \qquad (42)$$

and the functional integral becomes an Ising sum over the values $\varphi_x = \pm 1$. In the $n$-component case, the $\delta$ function $\delta(|\varphi_x|^2 - 1)$ fixing the relation $|\varphi_x|^2 = \sum_\alpha \varphi_{\alpha,x}^2$ appears and the $\sigma$ model is obtained instead of the Ising model [28].

The functional integrals are now functions of the single variable $t\kappa$ and the right-hand sides of Eqs. (9)–(12) depend only on this variable; this condition determines the renormalization-group functions of one variable $g$. The physical motivation of the limiting transition is based on the conditions

$$t \ll 1, \quad \kappa \gg 1, \quad t\kappa \gg 1, \qquad (43)$$

which ensure Eq. (38), but only two first conditions are really used when deriving Eq. (42);[7] i.e., it is valid under the conditions

$$t \ll 1, \quad \kappa \gg 1, \quad t\kappa \text{ is arbitrary}. \qquad (44)$$

Therefore, Eq. (42) can be used in the region $t\kappa \ll 1$, where gradient expansions are possible and the renormalized charge $g$ can be large. In view of Eqs. (39) and (40), the strong-coupling regime of the $\varphi^4$ theory corresponds to the limit

$$t \longrightarrow 0, \ tm_0^2 \longrightarrow -\infty, \ tm^2 \longrightarrow 0, \ m^2 \longrightarrow \infty, \quad (45)$$

which obviously does not coincide with Eqs. (37). The return to the parametric representation given by Eqs. (32) under the assumption

$$-tm_0^2 = \kappa \gg 1, \qquad (46)$$

provides the expressions

$$g = \left(\frac{n}{t\kappa}\right)^{d/2} g^*, \quad \beta(g) = \left(\frac{n}{t\kappa}\right)^{d/2} dg^*,$$
$$\eta_2(g) = \left.\frac{\beta_{\text{zero}}(-\kappa)}{g^* - g_{\text{zero}}(-\kappa)}\right|_{\kappa\to\infty} = -4, \qquad (47)$$

which give the asymptotic expressions

$$\beta(g) = dg, \quad \eta(g) = 0, \quad \eta_2(g) = -4 \ (g \longrightarrow \infty). \quad (48)$$

---

[7] The condition $t \ll 1$ is necessary for fluctuations of $|\varphi_x|^2$ in the gradient term to be insignificant.

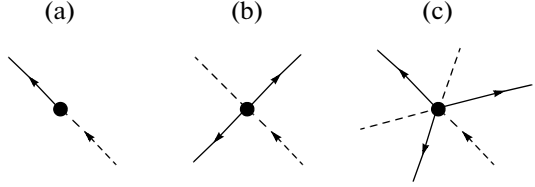

**Fig. 3.** Passage through a regular point $t_0$ under the conditions (a) $f'(t_0) \neq 0$; (b) $f'(t_0) = 0$, $f''(t_0) \neq 0$; and (c) $f'(t_0) = 0$, $f''(t_0) = 0$, and $f'''(t_0) \neq 0$. The solid and dashed lines show the directions of increasing and decreasing $f(t)$.

The last limit in Eqs. (47) is estimated taking into account that

$$g_{\text{zero}}(t) = g^* - \frac{2n}{n+2}\frac{1}{t^2},$$

$$\beta_{\text{zero}}(t) = -\frac{8n}{n+2}\frac{1}{t^2}, \quad t \longrightarrow \infty, \tag{49}$$

where the definitions of $\beta_{\text{zero}}(t)$ and $g_{\text{zero}}(t)$ and [1, Eqs. (30)] were used. It is interesting that the result for $\beta(g)$ coincides with Eq. (13); the difference between the results (13) and (30) for $\eta(g)$ and $\eta_2(g)$ is discussed in Section 7.

It is worth noting that representation (42) for the functional integrals can be used to calculate not only the renormalization-group functions, but also observables, which are obtained in the form

$$A_{\text{obs}} = \Lambda^{d_A} f_A(t\kappa), \tag{50}$$

where $d_A$ is the physical dimension of the observable $A_{\text{obs}}$. With the use of similar expressions for $g$ and $m$,

$$g = f_g(t\kappa), \quad m^2 = \Lambda^2 f_m(t\kappa), \tag{51}$$

Eq. (50) can be represented in the form

$$A_{\text{obs}} = m^{d_A} F(g), \tag{52}$$

which is free of the bare parameters $g_0$, $m_0$, and $\Lambda$; i.e., Eq. (52) presents the "renormalizability theorem" for the strong-coupling region.[8] It should be emphasized we do not take the continual limit in the bare theory and conserve the lattice as a convenient computational tool, excluding only the lattice constant $a$ from the physical results. The expansion of the exponential in Eq. (42) in the gradient term can provide the constructive expansions of Eq. (52) in negative powers of $g$.

---

[8] For the complete proof of renormalizability, it is necessary to analyze the possibility of excluding information on a particular form of the overlap integrals $J_x$ from the results such as Eq. (52). Note that in the classical renormalizability theorems [9, 22], independence of $\Lambda$ is proven not for quantities themselves (in the strong sense), but for the coefficients of their expansions in $g$ (in the weak sense). The strong and weak statements are equivalent to each other if there is a certain procedure of the summation of divergent series, which really exists [23]. However, to implement this procedure, it is necessary to analyze the analytic properties of Borel transforms and, in particular, to prove the absence [24] of renormalon singularities [25, 26]; this requires the renormalizability in the strong sense (see the comments at the end of work [27]). The equivalence of the strong and weak statements apparently solves the problem of the dependence on $J_x$, because such a dependence is absent in the expansion coefficients.

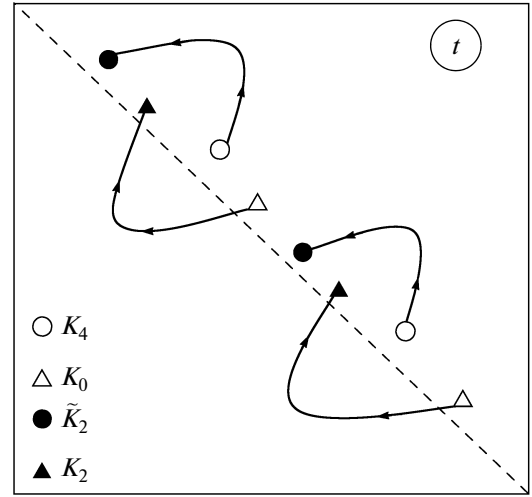

**Fig. 4.** Fragment of the complex $t$ plane along the ray $\arg t = \pm 3\pi/4$. The trajectories begin at the zeros of the integrals $K_0$ and $K_4$ and end at the zeros of $K_2$ and $\tilde{K}_2$.

## 5. TOPOLOGY OF THE TRAJECTORIES IN THE $t$ PLANE

Since the functions $\eta(g)$ and $\eta_2(g)$ in Eqs. (48) are different from the respective functions in Eqs. (13), it is interesting to determine the conditions of the applicability of these two results. This requires a more detail analysis of parametric representation (1).

Let the running parameter $t$ move along a certain continuous trajectory in the complex plane along which the relation $g = f(t)$ ensures the reality of the renormalized charge $g$ and its monotonic increase. If $f'(t_0) \neq 0$ at a certain point $t_0$ of the trajectory, there is one and only one direction of the passage through the point $t_0$ along which $g$ is real (see Fig. 3a): a trajectory reaching the point $t_0$ passes through it without a change in the direction. If $f'(t_0) = 0$ and $f''(t_0) \neq 0$, $g$ is real along two mutually perpendicular directions: there are two directions of increasing and two directions of decreasing $f(t)$ (see Fig. 3b). For this reason, a trajectory reaching the point $t_0$ should turn by an angle of $\pm 90°$. If $f'(t_0) = 0$, $f''(t_0) = 0$, and $f'''(t_0) \neq 0$, there are three directions of increasing and three directions of decreasing $f(t)$: a trajectory reaching the point $t_0$ can continue in the same direction or turn by an angle of $\pm 120°$ (see Fig. 3c), etc. It is easy to see that the trajectory cannot end at a regular point for the function $f(t)$, because a direction for its continuation always exists. It can end only at a singular point $t_c$ (finite or infinite) near which $g$ increases unboundedly remaining real.

According to Eq. (9), finite singular points $t_c$ are associated with the zeros of the integrals $K_2$ and $\tilde{K}_2$.

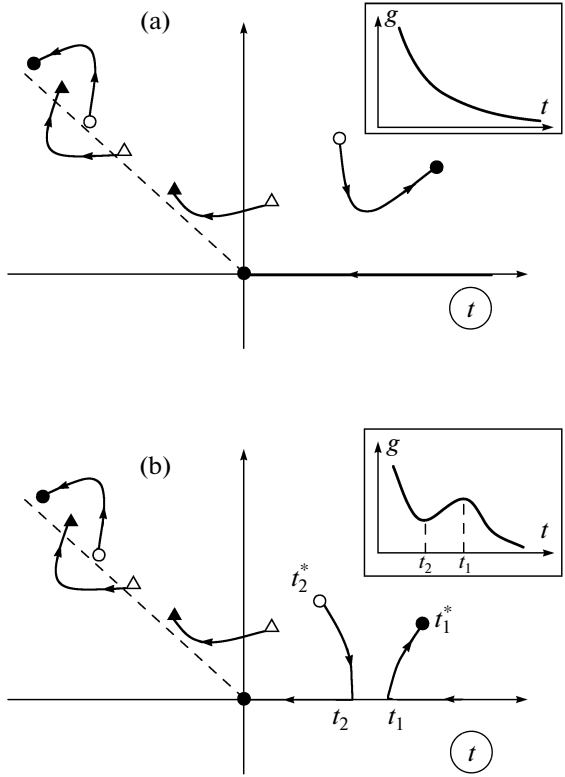

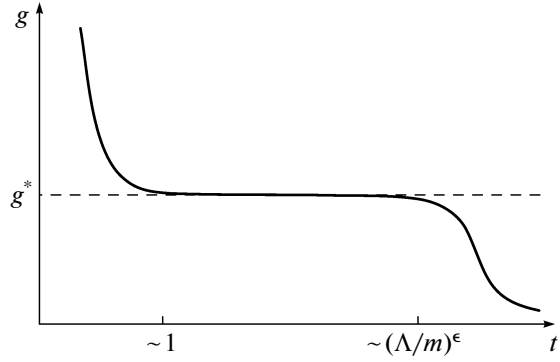

**Fig. 6.** Real function $g = f(t)$ on the real axis is close to a marginal dependence.

**Fig. 5.** Topology of the trajectories in the $t$ plane for the cases where the dependence $g = f(t)$ on the real axis is (a) monotonic and (b) nonmonotonic.

According to [1], each functional integral has the infinite number of zeros, which are located along the rays $\arg t = \pm 3\pi/4$ and concentrated at infinity. The variation of $g$ from 0 to $\infty$ corresponds to the infinite number of trajectories that begin at the zeros of $K_0$ or $K_4$ and end at the zeros of $K_2$ or $\tilde{K}_2$ (see Fig. 4). In small segments of the rays $\arg t = \pm 3\pi/4$, the zeros of the integrals are located quasiperiodically and it is reasonable to expect that all of them are closed to each other; only several zeros in the region $|t| \sim 1$, which can be topologically connected with trajectories passing through the real axis, can be "uncompensated."

Let us briefly discuss the possibility of going the trajectory to infinity. In the zero-dimensional case, it is possible to verify [1] that an increase in $\rho$ at $t = \rho e^{i\chi}$ is accompanied by a decrease in $g$ to zero at $\chi < 3\pi/4$ and by the saturation of $g$ at $\chi > 3\pi/4$; a similar situation is expected for the general case. At $d = 0$, the trajectory goes to infinity along the negative semiaxis, but then returns to the finite part of the complex plane; i.e., the trajectory passes through the point $t = \infty$, but does not end at it. The end of the trajectory at infinity (approached in another direction) is low probable, because for an unbounded increase in $g$, the trajectory should go to infinity, approaching one of the rays $\chi = \pm 3\pi/4$; in this case, the trajectory should pass near the infinite number of singular points and should not turn to any of them.

The trajectory that begins at $t = +\infty$ and passes along the real axis is of main interest. At $t \gg 1$, $g \propto g_0 = 1/t^2$, whereas at $t \ll 1$, $g \propto t^{-d/2}$; the latter dependence was discussed in Section 3. Two main scenarios are possible at intermediate $t$ values.

(i) If the function $g = f(t)$ is strictly monotonic on the real axis, then $f'(t) \neq 0$ and the trajectory cannot turn to the complex plane: it continues to $t = 0$, whereas all of the zeros of the functional integrals at complex $t$ values are closed to each other (see Fig. 5a).

(ii) If the function $g = f(t)$ is nonmonotonic on the real axis, the point of the maximum $t_1$ and the point of the minimum $t_2$ exist in the simplest case (see Fig. 5b). Then, the trajectory reaching the point $t = t_1$ from $t = +\infty$ should turn at this point at the right angle to the upper (or lower) half-plane and end at one of the singular points $t_1^*$. Similarly, the trajectory passing along the real axis to the point $t = 0$ should reach the real axis at the point $t_2$ and begin at a certain complex point $t_2^*$.

When case (i) is smoothly transformed to case (ii), the trajectories are reconnected (cf. Figs. 5a, 5b); in the marginal case $t_1 = t_2$ (corresponding to the saddle point), they are branched according to Fig. 3c. The real situation (Section 6) is close to the marginal case: the function $g = f(t)$ has a plateau from $t \sim 1$ to $t \sim (\Lambda/m)^\epsilon$ (see Fig. 6). It is reasonable to expect that change in lattice action (14) can "break" this function to both cases (i) and (ii). Therefore, the passage of the trajectory in the complex plane (with the implementation of results (13)) or along the real axis (with the implementation of results (48)) is determined by the method of the regularization of the functional integrals.

## 6. INFORMATION FROM THE GELL-MANN–LOW EQUATION

The Gell-Mann–Low equation makes it possible to analyze the relation of the renormalized charge $g$ at the scale $m$ to its value $g_\Lambda$ at the scale $\Lambda$; the latter value is directly related to the bare charge $g_0$, but does not coincide with it (see below); the relation $g_\Lambda \approx g_0$ is valid only in the weak-coupling region.

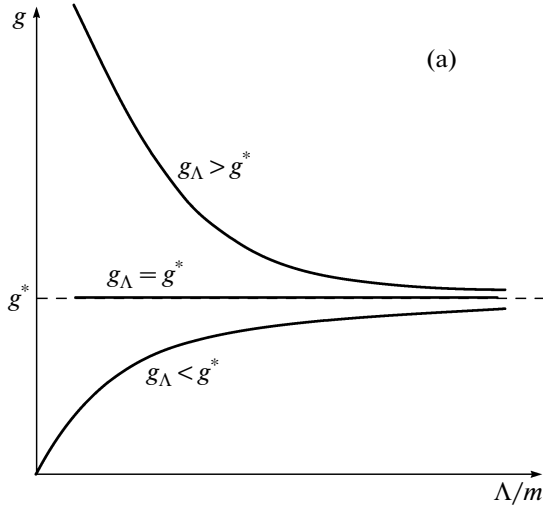

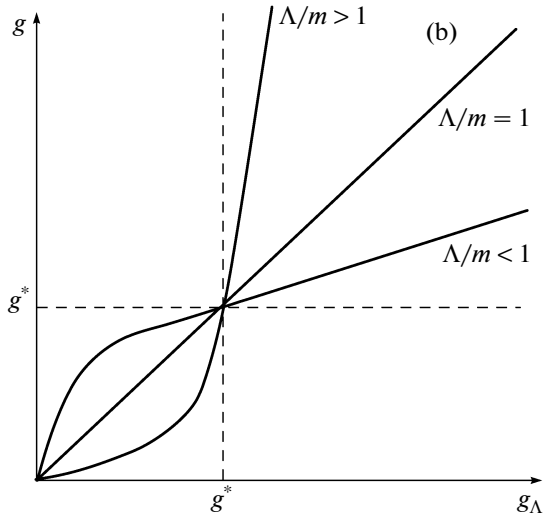

**Fig. 7.** Dependences of $g$ on (a) $\Lambda/m$ for various $g_\Lambda$ values and (b) $g_\Lambda$ for various $\Lambda/m$ values.

**1.** The Gell-Mann–Low equation at $d = 4 - \epsilon$ with small $\epsilon$ in the region $g \lesssim 1$ can be represented in the form

$$\frac{dg}{d\ln m} = \beta(g) = -\epsilon g + \beta_2 g^2 \qquad (53)$$

and has a stationary point $g^* = \epsilon/\beta_2$. The integration of Eq. (53) with the initial condition $g = g_\Lambda$ at $m = \Lambda$ provides

$$g = \frac{g_\Lambda (\Lambda/m)^\epsilon}{1 + \beta_2 g_\Lambda [(\Lambda/m)^\epsilon - 1]/\epsilon}, \quad g, g_\Lambda \lesssim 1. \qquad (54)$$

In the weak-coupling region, $g_\Lambda \approx g_0$ and, in the limit $\epsilon \longrightarrow 0$, the known result [29]

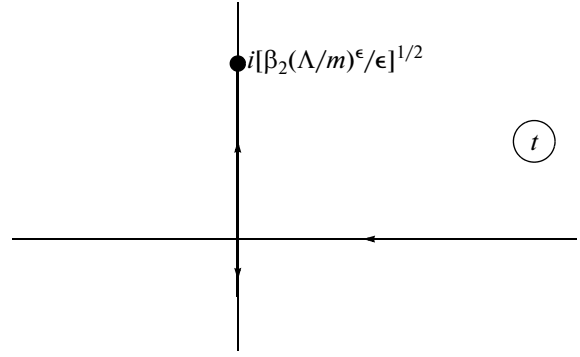

**Fig. 8.** Topology of the trajectory in the $t$ plane for $d = 4 - \epsilon$ with small $\epsilon$ values according to Eq. (59).

$$g = \frac{g_0}{1 + \beta_2 g_0 \ln(\Lambda/m)} \qquad (55)$$

is obtained. Equation (53) does not impose any restrictions on the relation between the scales $m$ and $\Lambda$. The dependence of $g$ on $\Lambda/m$ is increasing at $g_\Lambda < g^*$ and decreasing at $g_\Lambda > g^*$ and approaches the constant $g^*$ in the limit $\Lambda/m \longrightarrow \infty$ (see Fig. 7a). The dependences of $g$ on $g_\Lambda$ are shown in Fig. 7b; under the additional assumption

$$\beta(g) = \beta_\infty g, \quad g \gg 1, \qquad (56)$$

so that

$$g = \left(\frac{m}{\Lambda}\right)^{\beta_\infty} g_\Lambda, \quad g, g_\Lambda \gtrsim 1, \qquad (57)$$

it is clear that these dependences are monotonic[9] and have no pathologies.

However, if we suggest large $\Lambda/m$ and neglect unity in the square brackets in Eq. (54),

$$g = \frac{g_0 (\Lambda/m)^\epsilon}{1 + \beta_2 g_0 (\Lambda/m)^\epsilon/\epsilon}, \qquad (58)$$

the situation changes and large $g$ values are inaccessible for real $g_0$; setting $g_0 = 1/t^2$, parametric representation (1) is represented in the form

$$g = \frac{(\Lambda/m)^\epsilon}{t^2 + \beta_2 (\Lambda/m)^\epsilon/\epsilon},$$

$$\beta(g) = \frac{-\epsilon t^2 (\Lambda/m)^\epsilon}{[t^2 + \beta_2 (\Lambda/m)^\epsilon/\epsilon]^2}, \qquad (59)$$

where the trajectory in the $t$ plate turns at $t = 0$ at the right angle and ends at the Landau pole $i[\beta_2(\Lambda/m)^\epsilon/\epsilon]$ (see Fig. 8). Thus, the turn of the trajectory to the complex plane is not extraordinary and can easily appear under natural assumptions; note that this turn is not accompanied by any singularities of the β function. Equations (59) are valid only at $g, g_0 \lesssim 1$ and neither the position of the turning point $t = 0$ (where $g_0 \longrightarrow \infty$) nor the position of the pole (where $g \longrightarrow \infty$) is known with certainty.

---

[9] In the general case, this is clear from Fig. 7a, where the curves corresponding to different $g_\Lambda$ values do not intersect at finite $\Lambda/m$ values.

According to Eq. (58), $g$ has a regular expansion in the parameter $g_0(\Lambda/m)^\epsilon$, as in the diagrammatic calculations (see below), and, as a function of $g_0$, varies at a scale $\sim (m/\Lambda)^\epsilon \ll 1$ above which it approaches the constant $g^*$. This plateau holds to the scale $g_0 \sim 1$, where the differences of $g_0$ from $g_\Lambda$ appear (see Eq. (61)), the difference of the β function from Eq. (53) becomes significant, and the mass renormalization regime changes (see Eq. (35)); as a result, the transition to the strong-coupling regime discussed in Sections 3 and 4 occurs.

**2.** In the framework of the diagrammatic technique, the expansion of $g$ in $g_0$ has the structure[10]

$$g = \sum_{N=1}^{\infty} [g_0(\Lambda/m)^\epsilon]^N \sum_{K=0}^{N-1} A_N^K \times \left[\frac{1-(\Lambda/m)^{-\epsilon}}{\epsilon}\right]^K, \quad A_1^0 = 1. \quad (60)$$

At $m = \Lambda$, the relation of $g_\Lambda$ to $g_0$ is obtained in the form

$$g_\Lambda = g_0 + \sum_{N=2}^{\infty} A_N^0 g_0^N \equiv h(g_0), \quad (61)$$

so that the relation $g_\Lambda \approx g_0$ is valid only for small $g_0$. The dependence of $g$ on $g_\Lambda$ is monotonic (see Fig. 7b), whereas the character of the dependences of $g$ on $g_0$ is determined by the function $h(g_0)$. If the function $h(g_0)$ varies monotonically from zero to infinity with an increase in $g_0$, then the case shown in Fig. 5a occurs and the trajectory $t$ remains on the real axis. If the function $h(g_0)$ is nonmonotonic, the case shown in Fig. 5b occurs and the trajectories turn to the complex plane. If the function $h(g_0)$ is finite in the limit $g_0 \longrightarrow \infty$, the singularity at $t = 0$ disappears and large $g$ values are reached only at complex $t$. The function $h(g_0)$ is determined by the coefficients $A_N^0$, which can be varied by changing the lattice or using another regularization method.

The upper limit of the integration with respect to the momentum in the diagrammatic calculations at $d < 4$ is usually accepted as infinity; in this case, the term $(\Lambda/m)^{-\epsilon}$ in the square brackets in Eq. (60) can be omitted. As a result, Eq. (60) is represented in the form

$$g = \sum_{N=1}^{\infty} B_N [g_0(\Lambda/m)^\epsilon]^N,$$

$$B_N = \sum_{K=0}^{N-1} A_N^K \epsilon^{-K}, \quad (62)$$

and the relation of $g_\Lambda$ to $g_0$,

$$g_\Lambda = \sum_{N=1}^{\infty} B_N g_0^N, \quad B_1 = 1, \quad (63)$$

differs from Eq. (61). This difference is most pronounced for small $\epsilon$ values: according to Eq. (61) (which does not contain $\epsilon$), the relation $g_\Lambda \approx g_0$ is valid up to $g_0 \sim 1$, whereas Eq. (63) corresponds to the result $g_\Lambda = g_0/(1 + g_0/g^*)$ (following from Eq. (58)) and the dependence is saturated at $g_0 \sim \epsilon$. It should be emphasized that the transition from Eq. (61) to Eq. (63) does not involve any serious features: for $d < 4$ (after mass renormalization), the integrals converge at high momenta and it is physically insignificant whether the upper integration limit is taken as infinite or large but finite. Nevertheless, the relation of the renormalized charge to the bare charge changes drastically. This certainly does not mean the same changes in the observables, but indicates the possibility of two description methods.

(i) In the first description method that corresponds to Eq. (62) and is possible only at $d < 4$, the passage to the limit $\Lambda \longrightarrow \infty$ is performed at an early stage and the perturbation series in $g$ is constructed in the form that does not contain information on a cutoff. These series have real coefficients and provide real values of the observables with the usual summation methods [23]. The results independent of the cutoff cannot be matched with the lattice expansions discussed in Sections 3 and 4 and the passage of the trajectory $t$ to the complex plane looks natural.

(ii) In the second description method that corresponds to Eq. (60) and is the only possible method at $d = 4$ (i.e., in the actual field theory),[11] the explicit regularization method is implied. In this case, the renormalization-group functions, Green's functions, etc. after renormalization have finite limits in the limit $\Lambda \longrightarrow \infty$, but depend on details of cutoff (since the definition of the charge depends on these details); this dependence is expected to disappear only in the observables. In this description method, the trajectory $t$ can remain on the real axis at least for some regularization procedures.

---

[10] The $N$-loop approximation involves the integrations with respect to $N$ momenta; each integration gives the factor $k^{-\epsilon}$, which reduces to $\Lambda^{-\epsilon}$ and $m^{-\epsilon}$ in the upper and lower limits, respectively. For this reason, the $N$-loop contribution contains the factor $u_0^{N+1}$ multiplied by the $N$-order homogeneous polynomial composed of $\Lambda^{-\epsilon}$ and $m^{-\epsilon}$. Result (60) is obtained taking into account the relation $u_0 = g_0\Lambda^\epsilon$ after the appropriate grouping of the terms and the separation of the powers of $\epsilon$ from the coefficients for the correspondence in the limit $\epsilon \longrightarrow 0$ with the usual logarithmic expansion.

[11] It is also more natural in the condensed matter physics, where the cutoff certainly exists and the continual limit is possible only when it does not provide pathologies.

An example of small $\epsilon$ values clearly shows that the complex-valued bare charge $g_0$ is not attributed to any pathologies and has no physical meaning.

**3.** In the framework of expansion (62), the relation between $g$ and $g_0$ is completely determined by the $\beta$ function and can be analyzed in the general form. Integrating the Gell-Mann–Low equation,

$$\exp\{-\epsilon F(g)\} = \left(\frac{\Lambda}{m}\right)^\epsilon \exp\{-\epsilon F(g_\Lambda)\}, \quad (64)$$

$$F(g) = \int \frac{dg}{\beta(g)}$$

and setting

$$\beta(g) = \begin{cases} -\epsilon g, & g \longrightarrow 0 \\ \omega(g - g^*), & g \longrightarrow g^*, \end{cases} \quad (65)$$

it is easy to verify that the left-hand side of Eq. (64) is a regular function of $g$ and $g \approx (\Lambda/m)^\epsilon g_\Lambda \approx (\Lambda/m)^\epsilon g_0$ for small $g$ values. If the relation between $g_\Lambda$ and $g_0$ is defined so that

$$\exp\{-\epsilon F(g)\} = (\Lambda/m)^\epsilon g_0, \quad (66)$$

$$g_0 = \exp\{-\epsilon F(g_\Lambda)\},$$

then $g$ is a regular function of the parameter $g_0(\Lambda/m)^\epsilon$ as is required by expansion (62). At $g \longrightarrow g^*$, Eq. (66) provides

$$g^* - g = \left(\frac{\Lambda}{m}\right)^{-\omega} g_0^{-\omega/\epsilon} = \left(\frac{\Lambda}{m}\right)^{-\omega} t^{2\omega/\epsilon} \quad (67)$$

and the turn to the complex plane occurs at the point $t = 0$ at the angle

$$\chi = \frac{\pi}{2}\frac{\epsilon}{\omega}, \quad (68)$$

so that the complex $t$ values correspond to significantly complex (rather than simply negative) $g_0$ values. In this case, the entire picture is qualitatively the same as for small $\epsilon$: in particular, the dependences of $g$ on $g_0$ and $\Lambda/m$ are the same and correspond to the lower curve in Fig. 7a, it approaches the constant value $g^*$ in the limit $g_0 \longrightarrow \infty$, which is usually considered as the strong-coupling limit of the $\varphi^4$ theory [30].

Thus, in the renormalization scheme that is commonly accepted in the theory of phase transitions [12] and corresponds to expansion (62), the turn to the complex plane is inevitable and a singularity at $t = 0$ is absent. These properties does not contradict Section 3, because the definition of the bare charge in this scheme (see Eq. (63)) is different from that used in the lattice version of the functional integrals (see Eq. (61)).

## 7. SITUATION WITH ANOMALOUS DIMENSIONS

According to Section 6, the passage of the trajectory $t$ in the complex plane or on the real axis is determined by the renormalization scheme, i.e., is associated with the description method that has no physical meaning. The results for the asymptotic behavior of $\beta(g)$ confirm this point of view: they are the same for the singularities at $t = 0$ and at the complex point $t_c$ (cf. Eqs. (48) and (13)). Results (48) for the functions $\eta(g)$ and $\eta_2(g)$ are different from the respective results in Eqs. (13), but the anomalous dimensions have physical meanings only near the stationary point of the renormalization group $g^*$, whereas far from this point, they are technical constructions meaningful only in a particular scheme. Indeed, the functions $\eta(g)$ and $\eta_2(g)$ are determined by the $Z$ factors, which are physically meaningless by definition, because they are not involved in observables. The singular behavior of the $Z$ factors is manifested in observables only owing to a specific situation near the critical point. Let us illustrate this for the function $\eta(g)$.

As is known in the theory of phase transitions, the vertex $\Gamma_R^{(0,2)}$ (the inverse renormalized propagator) at $m^2 > 0$ has a regular expansion at small $p$ and a singular behavior at large $p$,

$$\Gamma_R^{(0,2)}(p)$$

$$= \begin{cases} m^2 + p^2 + \alpha_1 p^4 + \alpha_2 p^6 + \dots, & p \lesssim \xi^{-1}, \\ \sim p^{2-\eta}, & p \gtrsim \xi^{-1}. \end{cases} \quad (69)$$

The regular expansion with arbitrary coefficients $\alpha_1$, $\alpha_2$, ... satisfies the renormalization-group equations; hence, $\alpha_1 = \alpha_2 = \dots = 0$ can be taken and the regular solution $m^2 + p^2$ can be chosen at arbitrary $p$. The result $\Gamma_R^{(0,2)}(p) \sim p^{2-\eta}$ is the single solution only at $m^2 = 0$ (at the transition point) and its validity for $p \gtrsim \xi^{-1}$ at finite $m^2$ requires additional reasons: it does not contradict the regular expansion if the coefficients $\alpha_1$, $\alpha_2$, ... are appropriately chosen and ensure the asymptotic behavior $p^{2-\eta}$ at large $p$. For the case under consideration, $\eta = 0$ in one description method and indicates that the regular solution $m^2 + p^2$ exists at any $p$. However, this solution is always allowable and does not contradict the value $\eta = 2$ obtained in the other description method. Note that the instanton calculations by Polyakov [31] (see also [32]) can be treated [21] as the solution of the one-dimensional $\varphi^4$ theory in the limit $g_0 \longrightarrow \infty$; the result obtained for the pair correlation function,

$$G_2(x, y) = A \exp(-m|x-y|)$$

corresponds to $\Gamma_R^{(0,2)}(p) = m^2 + p^2$ in the momentum representation.

Let us give a more formal consideration based on the Callan–Symanzik equation[12]

$$\left[\frac{\partial}{\partial \ln \mu} + \beta(g)\frac{\partial}{\partial g} + \gamma_m(g)\frac{\partial}{\partial \ln \tau} - \eta(g)\right]\tilde{\Gamma}_R^{(0,2)} = 0, \quad (70)$$

where $\mu$ is the arbitrary momentum scale, $\tau \propto \delta m_0^2$ is the distance to the transition, and $\gamma_m(g) = 2 - \nu^{-1}(g) = \eta(g) - \eta_2(g)$. Its general solution for finite $p$ and $\tau$ can be represented in the form

$$\tilde{\Gamma}_R^{(0,2)}(p,\tau) = \mu^2 \exp\left\{\int dg \frac{\eta(g)-2}{\beta(g)}\right\}$$
$$\times F\left(\frac{p}{\mu}\exp\int\frac{dg}{\beta(g)}, \frac{\tau}{\mu^2}\exp\int\frac{dg}{\nu(g)\beta(g)}\right), \quad (71)$$

where $F(x, y)$ is an arbitrary function. Under the assumption that the functions $\eta(g)$ and $\nu(g)$ are constant in the $g$ range of interest, the expression

$$\tilde{\Gamma}_R^{(0,2)}(p,\tau) = \mu^2 A(g)^{\eta-2}$$
$$\times F\left(\frac{p}{\mu}A(g), \frac{\tau}{\mu^2}A(g)^{1/\nu}\right) \quad (72)$$

is obtained, where the function $A(g)$ in view of Eqs. (65) and (56) has the form

$$A(g) = \exp\int\frac{dg}{\beta(g)}$$
$$= \begin{cases} \sim g^{-1/\epsilon}, & g \longrightarrow 0 \\ \sim (g-g^*)^{1/\omega}, & g \longrightarrow g^* \\ \sim g^{1/\beta_\infty}, & g \longrightarrow \infty. \end{cases} \quad (73)$$

To obtain the finite solution in any of three limits (when $A(g)$ tends to zero or infinity), it should be constructed so that the dependence on $A(g)$ disappears. For $p = 0$ or $\tau = 0$, $F(0, y) \sim y^\alpha$ with $\alpha = \nu(2-\eta)$ and $F(x, 0) \sim x^\beta$ with $\beta = 2 - \eta$, respectively; thus, the known results

$$\tilde{\Gamma}_R^{(0,2)}(0,\tau) \sim \tau^{\nu(2-\eta)},$$
$$\tilde{\Gamma}_R^{(0,2)}(p,0) \sim p^{2-\eta} \quad (74)$$

---
[12] In the initial renormalization scheme corresponding to Eqs. (7) and (8), the Callan–Symanzik equation has a right-hand side [20, Sect. VI.A] and is inconvenient for the analysis. Here, only the possibility of the equivalence of the situations with $\eta = 0$ and 2 is demonstrated and a more convenient scheme [20, Sect. VI.C] is used where the tilde marks another method of the renormalization of $\Gamma^{(0,2)}$.

are obtained. In the general case, $F(x, y) \sim x^\beta y^\alpha$ with $\beta = \nu(2 - \eta - \alpha)$ can be accepted, so that

$$\tilde{\Gamma}_R^{(0,2)}(p,\tau) \sim \mu^2\left(\frac{\tau}{\mu^2}\right)^{\nu(2-\eta)}\left(\frac{p\tau^{-\nu}}{\mu^{1-2\nu}}\right)^\alpha \quad (75)$$

is a solution at arbitrary $\alpha$. In the general case, the solution can have the form of the superposition of functions of form (75); in particular, for the regular expansion in $p^2$,

$$\tilde{\Gamma}_R^{(0,2)}(p,\tau) = \mu^2\left(\frac{\tau}{\mu^2}\right)^{\nu(2-\eta)}\sum_{s=0}^{\infty} A_s\left(\frac{p\tau^{-\nu}}{\mu^{1-2\nu}}\right)^{2s}$$
$$= A_0\mu^2\left(\frac{\tau}{\mu^2}\right)^{\nu(2-\eta)} + A_1 p^2\left(\frac{\tau}{\mu^2}\right)^{-\nu\eta} \quad (76)$$
$$+ A_2\frac{p^4}{\mu^2}\left(\frac{\tau}{\mu^2}\right)^{-2\nu-\nu\eta} + \ldots.$$

With $\tilde{Z} = (\tau/\mu^2)^{\nu\eta}$, $m^2 = \mu^2(\tau/\mu^2)^{2\nu}$, and $A_0 = A_1 = 1$, Eq. (76) is represented in the form

$$\tilde{\Gamma}_R^{(0,2)}(p,\tau) = \tilde{Z}^{-1}\left(m^2 + p^2 + A_2\frac{p^4}{m^2} + \ldots\right), \quad (77)$$

i.e., the regular solution is possible for arbitrary $\eta$ value, as was mentioned above. Thus, the limits $\eta \longrightarrow 0$ and $\eta \longrightarrow 2$ can really correspond to the same physical situation.

The results for zero momenta such as $m^2 \sim \tau^{2\nu}$ and $\tilde{\Gamma}_R^{(0,2)}(0,\tau) \sim \tau^{\nu(2-\eta)}$ are constructive only near the critical point, when the relations of the observables to the renormalization-group charges can be linearized and the distance to the transition, $\tau$, is determined by the linear deviation of the controlling parameter from the critical value. Far from the transition point, such relations do not contain significant information, because the distance to the transition can be determined by different methods.

## 8. CONCLUDING REMARKS

The main reason against complex bare parameters is based on the representation of the $S$ matrix in terms of the Dyson $T$ exponential, according to which the bare Hamiltonian should be Hermitian for the unitarity of the theory.

The real situation is more complicated, as is clear from the Bogoliubov axiomatic construction of the $S$ matrix [9]. According to this construction, the most general form of the $S$ matrix is given by the $T$ exponential of $i\mathcal{A}$, where $\mathcal{A}$ is the sum of the bare action and the sequence of "integration constants," which are determined by quasi-local operators. In the regularized theory, the integration constants can be taken as zero and, thus, it is possible to return to the Dyson form of the $S$ matrix. However, in the process of renormalization, these constants are treated as nonzero and chosen from the condition of the absence of divergences; then, these constants are included into the action by changing its parameters. For this reason, the $S$ matrix for the truly continual theory is determined by the renormalized action, whereas the bare Hamiltonian and Schrödinger equation are ill-defined; from this point of view, the complex-valued bare parameters are of no significance.

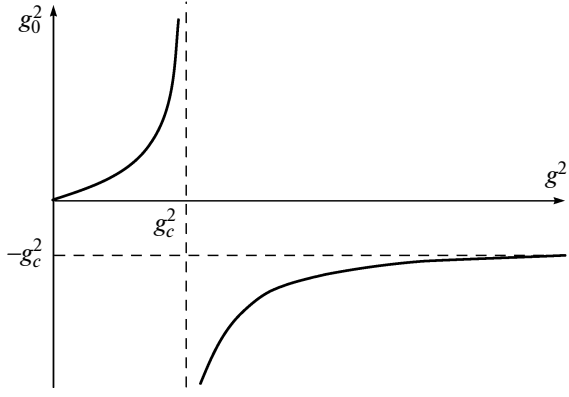

**Fig. 9.** Renormalized charge squared versus the bare charge squared in the Lee model.

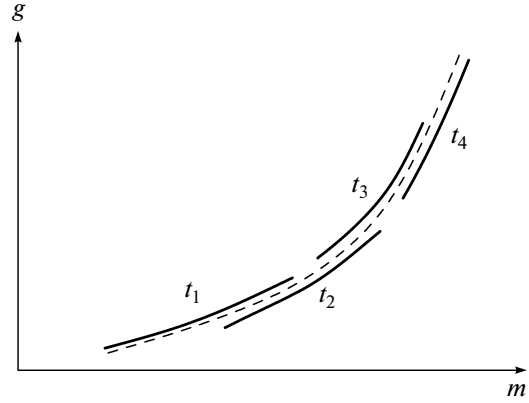

**Fig. 10.** Dependence $g \propto m^d$ for the singularity at the point $t = 0$ can be composed of pieces.

Some remaining questions concern the regularized theory, when both the bare and renormalized Lagrangians are meaningful and contradictory conclusions concerning the unitarity of the theory are possible. A similar situation was discussed for the exactly solvable Lee model [3] for which[13]

$$g^2 = \frac{g_0^2}{1 + g_0^2/g_c^2} \qquad (78)$$

and the bare charge $g_0$ is complex at large values of the renormalized charge $g$ (see Fig. 9). According to [4], the Lee model is physically unsatisfactory for $g > g_c$ owing to the presence of states with a negative norm (ghost states); however, Bender et al. [5, 6] recently showed that the problem of ghost states can be solved and the Lee model is an acceptable physical theory. The main idea of work [5] is that the analytic continuation of the parameters of the Hamiltonian to the complex plane should be accompanied by the modification of the scalar product for the corresponding Hilbert space,

$$(f, g) = \int f^*(x)g(x)dx \longrightarrow (f, g)_G = (f, \hat{G}g), \quad (79)$$

and the bare Hamiltonian with the appropriate choice of the operator $\hat{G}$ is Hermitian in terms of the new sca-lar product. As a result, all of the states of the Lee model have a positive norm and the $S$ matrix is unitary. A similar procedure should exist in the general case in order to eliminate the indicated contradiction.

---

[13] Cf. the Landau–Abrikosov–Khalatnikov formula for quantum electrodynamics [29].

The definition of the charge is ambiguous owing to the ambiguity of the renormalization scheme [11], which is associated with the arbitrariness of the integration constants in the Bogoliubov construction; consequently, the complex-valid bare charge $g_0$ has a relative meaning. The singularities of parametric representation (1) at $t = 0$ and at the complex point $t_c$ can be transformed to each other by redefining the bare charge $g_0$. For this reason, the result of this work for the $\beta$ function coincides with that obtained in [1] using absolutely different reasons.

Note that the result obtained for the asymptotic behavior of the $\beta$ function has a simple meaning. With the use of Eqs. (9) and (39), it is possible to write

$$g = -m^d \frac{K_4 K_0}{K_2^2}, \quad m^2 = \frac{K_2}{\tilde{K}_2}. \qquad (80)$$

For the dependence $g \propto m^d$, the result $\beta(g) = dg$ trivially follows from definition (8) for the $\beta$ function. Its validity in the asymptotic region is ensured under the following conditions: (i) the limit $m \longrightarrow \infty$ can be reached for the constant ratio $K_4 K_0 / K_2^2$ and (ii) this limit is possible owing to change only in $m_0$ (with unchanged $g_0$ and $\Lambda$ values). These conditions are easily satisfied near the zero of $\tilde{K}_2$ in the complex plane. For the singularity at $t = 0$, the indicated conditions are not satisfied, but their weakened forms are valid: change in $m_0$ can ensure change in $m$ under the condition $K_4 K_0 / K_2^2 = $ const in a certain wide range and this range can be shifted towards larger values by varying $t$ (see Section 4). Thus, the dependence $g \propto m^d$ can be composed from pieces by choosing a certain decreasing sequence $t_1 > t_2 > t_3...$ (see Fig. 10).

The objection is possible against the proposed scheme due to the fact that functional integrals are used in the "unphysical" regime $\xi \lesssim a$. However (from the conservative point of view), in principle, one can object to the lattice interpretation of the functional integrals, as well as to any other regularization method.[14] The main reason in favor of the applicability of such an approach is the possibility of removing all of the attributes of the bare theory from the physical results, but this reason can also be applied to the proposed scheme, at least concerning the results given by Eqs. (48). The bare theory is an auxiliary construction and any "physical" requirements to it are redundant. At the same time, conditions (44) can be strengthened and the passage to the following limit can be performed:

$$t \longrightarrow 0, \quad \kappa \longrightarrow \infty, t\kappa = \text{const}. \tag{81}$$

In this case, the passage to Eq. (42) does not require any approximations, but Eq. (42) holds the strict equivalence with the $\varphi^4$ theory under a special choice of the bare parameters, which ensures the conservation of the form of the Lagrangian in the process of renormalizations. At $t\kappa \gg 1$, result (42) satisfies all of the physical requirements and corresponds to the evidently correct theory; at $t\kappa \lesssim 1$, this result is a strict analytic continuation of this theory. Finally, in the condensed matter physics, the lattice Hamiltonian is an admissible microscopic Hamiltonian and can be used in any regime. Correspondingly, results (48) are certainly valid in the theory of phase transitions.

Let us briefly discuss the dependence of the results on the configuration of the overlap integrals $J_x$. In fact, the exclusion of $J_x$ in the general form is not necessary; these integrals can rather be chosen so that the lattice spectrum $\epsilon(p)$ is maximally close to the square spectrum. The result of such a procedure is known and corresponds to the approximation of almost free electrons in the theory of solids. In this case, the "empty lattice" limit rather than the continual limit $a \longrightarrow 0$ corresponds to the correct field theory.

To conclude, it is worth noting that the upper estimate ("triviality bound") for the mass of the Higgs boson based on the triviality of the $\varphi^4$ theory [33] is groundless. In the case of the asymptotic behavior $\beta(g) \sim g$, the Landau pole is absent and internal limitations on the applicability of the Standard Model implied in this estimate really do not exist.

---

[14] Such ideas arise, e.g., in relation to the concept of renormalons [25, 26]: the renormalon singularities are caused by arbitrarily high momenta and are removed by any regularization procedure.